\documentclass[useAMS,usenatbib]{mn2e}
\usepackage{graphicx}

\title{High resolution infrared spectra of NGC~6342 and NGC~6528:
two moderately reddened Bulge Globular Clusters
\thanks{Data presented herein were obtained at the W.M.Keck Observatory,
        which is operated as a scientific partnership among the California
        Institute of Technology, the University of California, and the National
        Aeronautics and Space Administration. The Observatory was made possible
        by the generous financial support of the W.M. Keck Foundation.}}

\author[Origlia, Valenti \& Rich]{L. Origlia$^1$, E. Valenti$^{2,1,3}$,
        R.~M. Rich$^3$  \\
 $^1$ INAF-Osservatorio Astronomico di Bologna, Via Ranzani 1, I-40127 Bologna,
      Italy, e-mail livia.origlia@bo.astro.it \\
 $^2$ Dipartimento Astronomia, Universit\`a di Bologna,  
      Via Ranzani 1, I-40127 Bologna, Italy, e-mail 
      elena.valenti3@unibo.it \\
      $^3$ Department of Physics and Astronomy, University of California
      at Los Angeles, Los Angeles, CA 90095-1562, e-mail rmr@astro.ucla.edu \\
       }
\date{\today}

\begin{document}
\pagerange{\pageref{firstpage}--\pageref{lastpage}} \pubyear{2004}
\maketitle
\label{firstpage}

\begin{abstract}
Using the NIRSPEC spectrograph at Keck II, we have obtained
infrared echelle spectra covering the range $1.5-1.8~\mu \rm m$ for
the moderately reddened bulge globular clusters NGC~6342 and
NGC~6528, finding [Fe/H]=--0.60 and --0.17~dex, respectively.
We measure an average $\alpha$-enhancement of $\approx+0.33$~dex
in both clusters,
consistent with previous measurements on other metal rich bulge clusters, 
and favoring the scenario of a rapid bulge formation and chemical enrichment.
We also measure very low  $\rm ^{12}C/^{13}C$ isotopic ratios 
($\approx$5 in NGC~6342 and $\approx$8 in NGC~6528),
suggesting that extra-mixing mechanisms due to {\it cool bottom processing}
are at work during evolution along the Red Giant Branch.

\end{abstract}

\begin{figure*}
\centering
\includegraphics[]{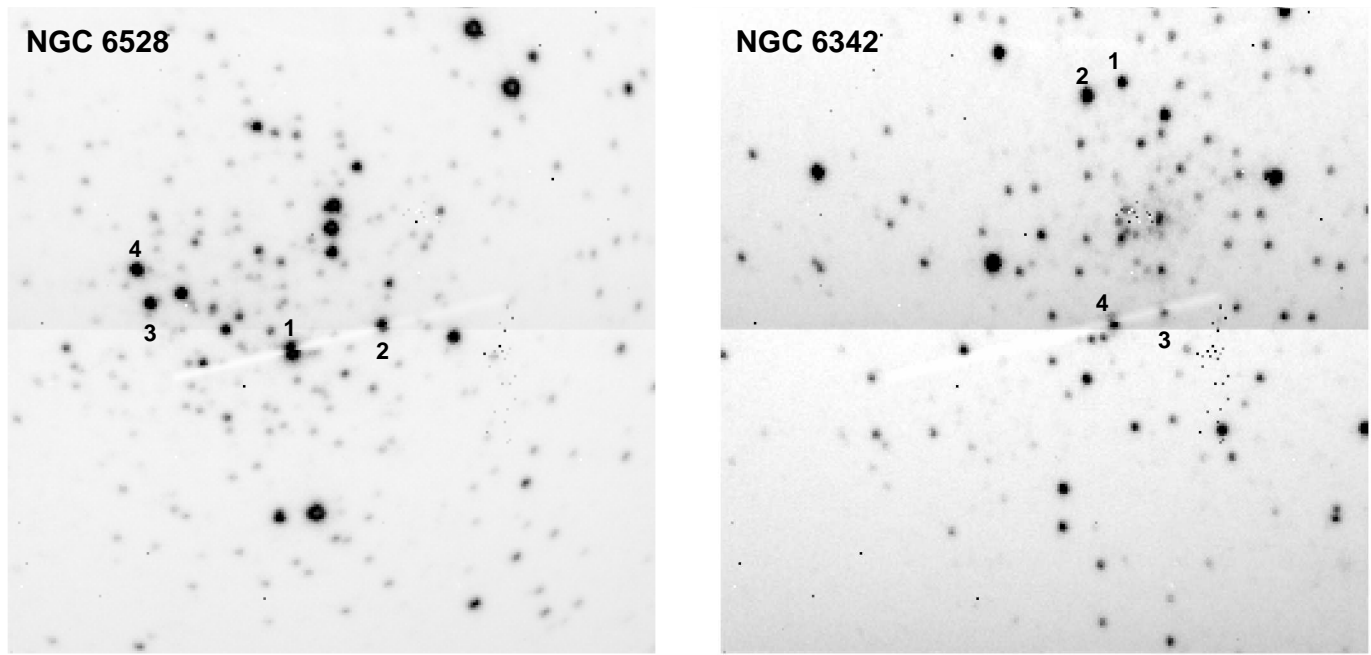}
\caption{
H band images of the core regions of NGC~6528 and NGC~6342 as
imaged by the slit viewing camera (SCAM) of NIRSPEC.
The field of view is 46\arcsec on a side
and the image scale is
$0\farcs183$$~pixel^{-1}$; the slit is 24\arcsec\ long.
The observed stars are numbered (cf. Table~1).
}
\label{fig1}
\end{figure*}
 
\section{Introduction}

Infrared, high resolution spectroscopy has tremendous potential as
a powerful tool to study both distant and obscured stellar populations.
For the heavily obscured regions of the Galactic bulge and center,
infrared spectroscopy offers the best approach to measuring the 
composition of the old stellar populations.

Over the past few years we have commenced
a high-resolution spectroscopic survey of the Galactic 
bulge in the near-IR using NIRSPEC, a high throughput infrared (IR) echelle
spectrograph at the Keck Observatory \citep{ml98}.
H--band (1.5--1.8 $\mu$m) spectra
of bright giants in the bulge globular clusters and field population are ideal
for detailed abundance analysis of Fe, 
C, O and other $\alpha $-elements, using the approach of synthesizing the
entire spectrum.  By observing the most luminous giants in the infrared,
we have the possibility of placing the obscured stars and globular clusters
toward the galactic center, and populations well studied in the optical,
on the same abundance scale.
The abundance distributions in the {\it cluster} and {\it field}
populations are important in constraining the history of bulge formation and 
chemical enrichment \citep{mw97}.

We have used this method to derive abundances for
four bulge globular clusters: the results for NGC~6553 
and Liller~1 are given in \citet{ori02}, while abundances for 
Terzan~4 and Terzan~5 are reported in \citet{ori04}.
We find $\alpha$-enhancement at a level of a factor between 2 and 3 over 
the whole range of metallicity spanned by the clusters in our survey, 
from [Fe/H]$\approx$--1.6 (cf. Terzan~4) up to [Fe/H]$\approx$--0.2 (cf. Terzan~
5).
In this paper we present the high resolution
IR spectra and the abundance analysis of  
four bright giants in NGC~6342 and NGC~6528,
two bulge globular clusters with relatively low reddening.

The most recent metallicity estimates based on 
optical \citep{hr99} and near-IR photometry \citep{mom03,val04} report
global metallicities [M/H]$\approx-0.4$ and $\approx+0.0$ dex and iron abundance
s 
[Fe/H]$\approx-0.6$ and $\approx-0.2$ for NGC~6342 
and NGC~6528, respectively.
NGC~6342 has never been studied at high spectral resolution before, 
while NGC~6528 has been observed with HIRES at Keck I by 
\citet{car01}. They found iron abundance slightly in excess of solar 
([Fe/H]=+0.07) and $\alpha$-enhancement of Ca ([Ca/Fe]=+0.2) and Si 
([Si/Fe]=+0.4) and marginal enhancement of O, Mg and Ti.
Very recently, \citet{zoc04} observed NGC~6528 giants with UVES at VLT,
finding [Fe/H]=--0.15 and moderate $\alpha$-enhancement of O, Si and Mg by
$\approx$+0.1--0.2 dex. 

Our observations and data reduction follow in Sect.~2.
Sect.~3 discusses our abundance analysis and
in Sect.~4 the resulting metallicities and radial velocities are presented.
We discuss our findings in Sect.~5.

\section{Observations and Data Reduction}

Near infrared, high-resolution echelle spectra of four bright giants
in the core of the bulge globular
clusters NGC~6342 and NGC~6528 have been acquired
during two observational campaigns in July 2002 and 2003.
We used the infrared spectrograph NIRSPEC \citep{ml98}
which is at the Nasmyth focus of the Keck~II telescope.
The high resolution echelle mode, with a slit width of $0\farcs43$
(3 pixels) and a length of 24\arcsec\
and the standard NIRSPEC-5 setting, which
covers most of the 1.5--1.8 micron H-band,
have been selected.  Typical exposure times (on source)
ranged from 8 to 16 minutes.
Fig.~\ref{fig1} shows the H band images of the observed core region of
NGC~6342 and NGC~6528
taken with the slit viewing camera
(SCAM) of NIRSPEC, which
has a field of view of 46\arcsec$\times$46\arcsec\
and a scale of $0\farcs183$$~pixel^{-1}$.


The raw two dimensional spectra were processed using the
REDSPEC IDL-based package written
at the UCLA IR Laboratory.
Each order has been
sky subtracted by using the pairs of spectra taken
with the object nodded along the slit, and subsequently
flat-field corrected.
Wavelength calibration has been performed using arc lamps and a second order
polynomial solution, while telluric features have been removed by
dividing by the featureless spectrum of an O star.
At the NIRSPEC resolution of R=25,000 several single
roto-vibrational OH lines and
CO bandheads can be measured to derive accurate oxygen and carbon abundances.
Other metal abundances can be derived from the atomic lines
of Fe~I, Mg~I, Si~I, Ti~I and Ca~I.
Abundance analysis is performed by using full spectral synthesis
techniques and equivalent width measurements of representative lines.

\section{Abundance Analysis}

We compute suitable synthetic spectra
of giant stars by varying the stellar parameters and the
element abundances using an updated
version of the code described in \citet{ori93}.
The main characteristics of the code have been widely discussed in our previous 
papers 
\citep{ori02,ori04} and they will not be repeated here.
The code 
uses the LTE approximation and is based
on the molecular blanketed model atmospheres of
\citet{jbk80} at temperatures $\le$4000~K
and the ATLAS9 models for temperatures above 4000~K.
The reference solar abundances are from
\citet{gv98}.

Photometric estimates of the stellar parameters are initially used 
as input to produce a grid of model spectra, 
allowing the abundances 
and abundance patterns to vary over a large range and the stellar parameters 
around the photometric values.  
The model which better reproduces the overall observed spectrum and 
the equivalent widths of selected lines is chosen as the best fit model.
We measure equivalent widths in the observed spectrum (see Table~\ref{tab1}), in

the best fit model and in four additional models which are, respectively, $\pm$0
.1 and
$\pm$0.2 dex away from the best-fitting.
 This approach gives us the uncertainties
listed in Table~\ref{tab2}.

Stellar parameter uncertainty of $\pm$200~K in temperature (T$_{eff}$), $\pm$0.5
~dex in 
log-gravity (log~g) and $\pm$0.5~km~s$^{-1}$ in microturbulence velocity ($\xi$)
, 
can introduce a further systematic $\le$0.2~dex uncertainty in the 
absolute abundances.
However, since the CO and OH molecular line profiles are very sensitive to 
effective temperature, gravity, and microturbulence variations, 
they constrain better the values of these parameters,  
significantly reducing their initial range of variation and
ensuring a good self-consistency of the overall spectral
synthesis procedure \citep{ori02,ori04}.
Solutions with  
$\Delta $T$_{\rm eff}$$=\pm$200~K, $\Delta $log~g=$\pm$0.5~dex and
$\Delta \xi$$=\mp$0.5~km~s$^{-1}$ and corresponding 
$\pm$0.2~dex abundance variations from the best-fitting one are indeed less
statistically
significant (typically at $1\le\sigma\le3$ level only, \citet{ori04}). 
Moreover, since the
stellar features under consideration show a similar trend
with variation in the stellar parameters, although with different
sensitivity, {\it relative } abundances are less
dependent on stellar parameter assumptions, 
reducing the systematic uncertainty 
to $<$0.1~dex.
\begin{table*}
\begin{center}
\caption[]{$\rm (J-K)_0$ colors, heliocentric radial velocity and
equivalent widths (m\AA) of some representative lines
for the observed stars in NGC~6342 and NGC~6528.}
\label{tab1}
\begin{tabular}{lcccccccccc}
\hline\hline
 & & & & & & & & & & \\
 & & NGC~6342 & & & & & &NGC~6528 & & \\
 & & & & & & & & & & \\
\hline
 & & & & & & & & & & \\
star                & \#1 & \#2 & \#3& \#4 &&&\#1  & \#2 & \#3  & \#4 \\
ref \#$^a$          & 53 & 16 & 48  &19 &&& 3139&3101 & 3167 &3166 \\
$\rm (J-K)_0^b$     &0.64 &0.80 &0.67& 0.77&&& 1.01&0.94 & 0.77 &0.78\\
$\rm M_{bol}$     &--0.6 &--1.4 &--0.6& --1.4&&&--3.7 &--2.3 & --2.4 &--2.6\\
$v_r$ [km~s$^{-1}$] &+111 &+111 &+111& +121&&&+215 &+215 & +200 &+210 \\
Ca~$\lambda $1.61508& 80  &132  & 80 & 113 &&& 279 & 261 &  231 & 209 \\
Fe~$\lambda $1.61532& 158 &186  & 145& 164 &&& 250 & 265 &  257 & 247 \\
Fe~$\lambda $1.55317& 145 &168  & 143& 153 &&& 213 & 227 &  230 & 218 \\
Mg~$\lambda $1.57658& 396 &408  &396 & 377 &&& 439 & 447 &  462 & 435\\
Si~$\lambda $1.58884& 399 &458  &400 &  397&&& 534 & 518 &  514 & 503 \\
OH~$\lambda $1.55688&  62 &211  & 62 &  136&&& 358 & 320 &  306 & 280 \\
OH~$\lambda $1.55721&  63 &228  & 62 &  137&&& 367 & 324 &  308 & 285\\
Ti~$\lambda $1.55437& 187 &290  & 187&  264&&& 413 & 418 &  397 & 429 \\
\hline
\end{tabular}
\end{center}
$^a$ Stars in NGC~6342 from \citet{val04},
in NGC~6528 from \citet{fer00}.\\
$^b$ Reddening corrected colors adopting E(B-V)=0.57 for NGC~6342 and
E(B-V)=0.62 for NGC~6528 (see Sect.~4).
\end{table*} 

\begin{table*}
\begin{center}
\caption[]{Adopted stellar atmosphere parameters and abundance estimates.}
\label{tab2}
\begin{tabular}{lcccccccccc}
\hline\hline
 & & & & & & & & & & \\
 & & NGC~6342 & & & & & &NGC~6528 & & \\
 & & & & & & & & & & \\
\hline
 & & & & & & & & & & \\
star            & \#1  & \#2  & \#3  & \#4  &&\#1  & \#2  & \#3  & \#4 \\
T$_{\rm eff}$ [K] & 4250 & 4000 &4250  &4000  && 3600 & 3800 & 4000 & 4000\\ 
log~g         & 1.5  & 1.0  &1.5   &1.0   && 0.5  & 0.5  & 1.0  & 1.0\\
$\xi$ [km~s$^{-1}$]   & 2.0  & 2.0  & 2.0  & 2.0  && 2.0  & 2.0  & 2.0& 2.0\\
$\rm [Fe/H]$   &--0.61&--0.57&--0.59&--0.62&&--0.21&--0.16&--0.15&--0.17\\
               &$\pm$.09 &$\pm$.08 &$\pm$.09 &$\pm$.09 &&$\pm$.06 &$\pm
$.07 &$\pm$.08 &$\pm$.08 \\
$\rm [O/Fe]$        & +0.33& +0.30& +0.29& +0.30&& +0.32& +0.27& +0.38& +0.33\\
                   &$\pm$.11 &$\pm$.09 &$\pm$.10 &$\pm$.10 &&$\pm$.10 &$\pm
$.09 &$\pm$.08 &$\pm$.09 \\
$\rm [Ca/Fe]$     & +0.40 & +0.37& +0.39& +0.37&& +0.41& +0.36& +0.35& +0.37\\
                 &$\pm$.17 &$\pm$.14 &$\pm$.17 &$\pm$.17 &&$\pm$.11 &$\pm
$.11 &$\pm$.12 &$\pm$.13 \\
$\rm [Si/Fe]$      & +0.36& +0.40& +0.39& +0.32&& +0.31& +0.26& +0.35& +0.27\\
                  &$\pm$.18 &$\pm$.17 &$\pm$.18 &$\pm$.18 &&$\pm$.19 &$\pm
$.19 &$\pm$.19 &$\pm$.19 \\
$\rm [Mg/Fe]$     & +0.37& +0.37& +0.39& +0.37&& +0.31& +0.36& +0.37& +0.37\\
                  &$\pm$.15 &$\pm$.16 &$\pm$.15 &$\pm$.15 &&$\pm$.14 &$\pm
$.14 &$\pm$.14 &$\pm$.14 \\
$\rm [Ti/Fe]$      & +0.21& +0.27& +0.24& +0.27&& +0.22& +0.26& +0.37& +0.37\\
                    &$\pm$.19 &$\pm$.19 &$\pm$.19 &$\pm$.19 &&$\pm$.13 &$\pm
$.13 &$\pm$.13 &$\pm$.14 \\
$\rm [\alpha/Fe]^a$  & +0.33& +0.35& +0.35& +0.34&& +0.31& +0.31& +0.36& +0.34\\
                    &$\pm$.13 &$\pm$.12 &$\pm$.13 &$\pm$.13 &&$\pm$.10 &$\pm
$.10 &$\pm$.11 &$\pm$.11 \\
$\rm [C/Fe]$      &--0.34&--0.33&--0.31&--0.37&&--0.29&--0.34&--0.25&--0.53\\
                 &$\pm$.11 &$\pm$.11 &$\pm$.11 &$\pm$.11 &&$\pm$.09 &$\pm
$.10 &$\pm$.10 &$\pm$.11 \\
\hline
\end{tabular}
\end{center}
$^a$ $\rm [\alpha/Fe]$ is the average $\rm [<Ca,Si,Mg,Ti>/Fe]$ abundance ratio
(see Sect.~4).
\end{table*}

\section{Results}
\label{results}
By combining full spectral synthesis analysis with equivalent width
measurements, we derive abundances of Fe, C, O
and $^{12}$C/$^{13}$C for the four observed
giants in NGC~6342 and NGC~6528.
The abundances of additional $\alpha-$elements Ca, Si, Mg and Ti are obtained
by measuring a few major atomic lines.

The near-IR spectra of cool stars also contain many CN molecular lines. 
However, at the NIRSPEC resolution most of them are very faint and blended 
with the stronger CO, OH and atomic lines.
By performing full spectral synthesis and analyzing the few lines 
reasonably detectable (although contaminated), 
one can obtain rough estimates of nitrogen abundances.

Stellar temperatures (see Table~\ref{tab2}) are both estimated from the $\rm (J-
K)_0$ 
colors (see Table~\ref{tab1}) and molecular lines,
gravity from theoretical evolutionary tracks,
according to the location of the stars on the Red Giant Branch (RGB), 
and adopting an average microturbulence velocity of 2.0 km/s
\citep[see also][]{ori97}.
Equivalent widths (see Table~\ref{tab1}) are computed by Gaussian fitting 
the line profiles and the overall uncertainty is $\le$10\%.

In order to check further the statistical significance of our best-fitting
 solution,
we compute synthetic spectra with 
$\Delta $T$_{\rm eff}$$=\pm$200~K, $\Delta $log~g=$\pm$0.5~dex and
$\Delta \xi$$=\mp$0.5~km~s$^{-1}$, and with corresponding simultaneous 
variations 
of $\pm$0.2~dex of the C and O abundances to reproduce the depth of the
molecular features.
We follow the strategy illustrated in \citet{ori04}.
As a figure of merit we adopt
the difference between the model and the observed spectrum (hereafter $\delta$).

In order to quantify systematic discrepancies, this parameter is
more powerful than the classical $\chi ^2$ test, which is instead
equally sensitive to {\em random} and {\em systematic} scatters.
 
Since $\delta$ is expected to follow a Gaussian distribution,
we compute $\overline{\delta}$ and the corresponding standard deviation
for our best-fitting solution and the other models 
with the stellar parameter and abundance variations quoted above.
We then extract 10,000 random subsamples from each
{\it test model} (assuming a Gaussian distribution)
and we compute the probability $P$
that a random realization of the data-points around
a {\it test model} display a $\overline{\delta}$ that is compatible
with an ideal best-fitting model with a $\overline{\delta}$=0. 
$P\simeq 1$ indicates that the model is a good representation of the
observed spectrum.
The statistical tests are performed on portions of the spectra
mainly containing the CO bandheads and the OH lines which are the 
most sensitive to the stellar parameters.

\subsection{NGC~6342}
\label{n6342}

In order to obtain a photometric estimate of the stellar temperatures
and the bolometric magnitudes we use the near IR photometry by \citet{val04} 
and their E(B-V)=0.57 reddening and (m-M)$_0$=14.63 distance modulus. 
We also use the color-temperature transformations and
bolometric corrections of \citet{mon98}, specifically
calibrated for globular cluster giants.
We constrain effective temperatures in the range 4000--4500~K, and
we estimate bolometric magnitudes 
$\rm M_{bol}\approx-0.6$ for stars \#1 e 3 and $\rm M_{bol}\approx-1.4$ 
for stars \#2 and 4 (see Table~\ref{tab1}).
The final adopted temperatures, obtained by best-fitting the CO and in 
particular the OH molecular bands which are 
especially temperature sensitive in cool giants, are reported in
 Table~\ref{tab2}.

\begin{figure*}
\centering
\includegraphics[width=18cm]{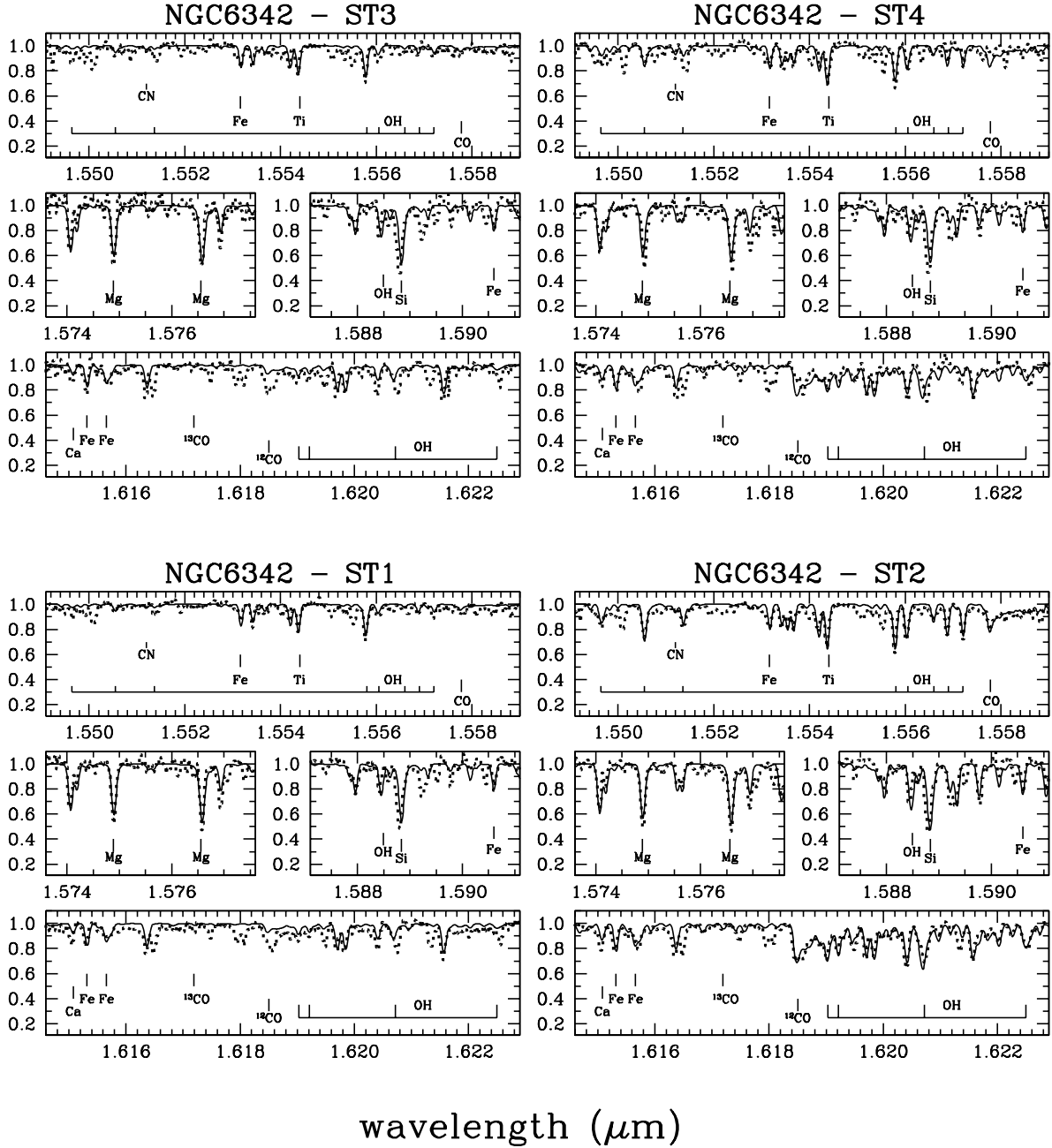}
\caption{
Selected portions of the observed echelle spectra (dotted lines) of the
four giants in NGC~6342 with our best-fitting synthetic spectrum
(solid line) superimposed. A few important molecular and atomic lines
of slightly interest are marked.
}
\label{fig2}
\end{figure*}


Fig.~\ref{fig2} shows
our synthetic best fits superimposed on the
observed spectra of the four giants in NGC~6342.
From our overall spectral analysis 
we find average [Fe/H]$=-0.60\pm0.01$, [O/Fe]$=0.31\pm 0.01$ and 
$\rm [\alpha/Fe]=0.34\pm0.01$
(see Table~\ref{tab2}).
We also measure an average carbon depletion ([C/Fe]=--0.34$\pm0.01$~dex)
and low $\rm ^{12}C/^{13}C\approx 5\pm 1$
isotopic ratio and enhancement (by a factor of 2-3) of [N/Fe].
As shown in Fig.~\ref{fig3} our best-fitting solutions  
have an average probability $\-P>$0.99 to be statistically 
representative of the observed spectra. 
The other {\it test models} with different assumptions for the stellar 
parameters are only significant at $>1.5~\sigma$ level. 

From the NIRSPEC spectra we also derived stellar heliocentric radial velocities
(see Table~\ref{tab1}), finding   an average value of
$+114\pm $3~km/s with a dispersion $\sigma=5$~km/s,  
in good agreement with the one obtained 
by \citet{dub97} who find $\rm v_r=+118\pm2$~km/s and $\sigma$=5~km/s.

\subsection{NGC~6528}
\label{n6528}

We use the near IR photometry of \citet{fer00}  
and their E(B-V)=0.62 reddening and (m-M)$_0$=14.37 distance modulus. 
We find photometric temperatures in the 3600-4000~K range and bolometric 
magnitudes between -2.3 and -3.7 (see Table~\ref{tab1}).
The final adopted temperatures, 
obtained by best-fitting the CO and the OH molecular bands, 
are reported in Table~\ref{tab2}.
\begin{figure*}
\centering
\includegraphics[width=18cm]{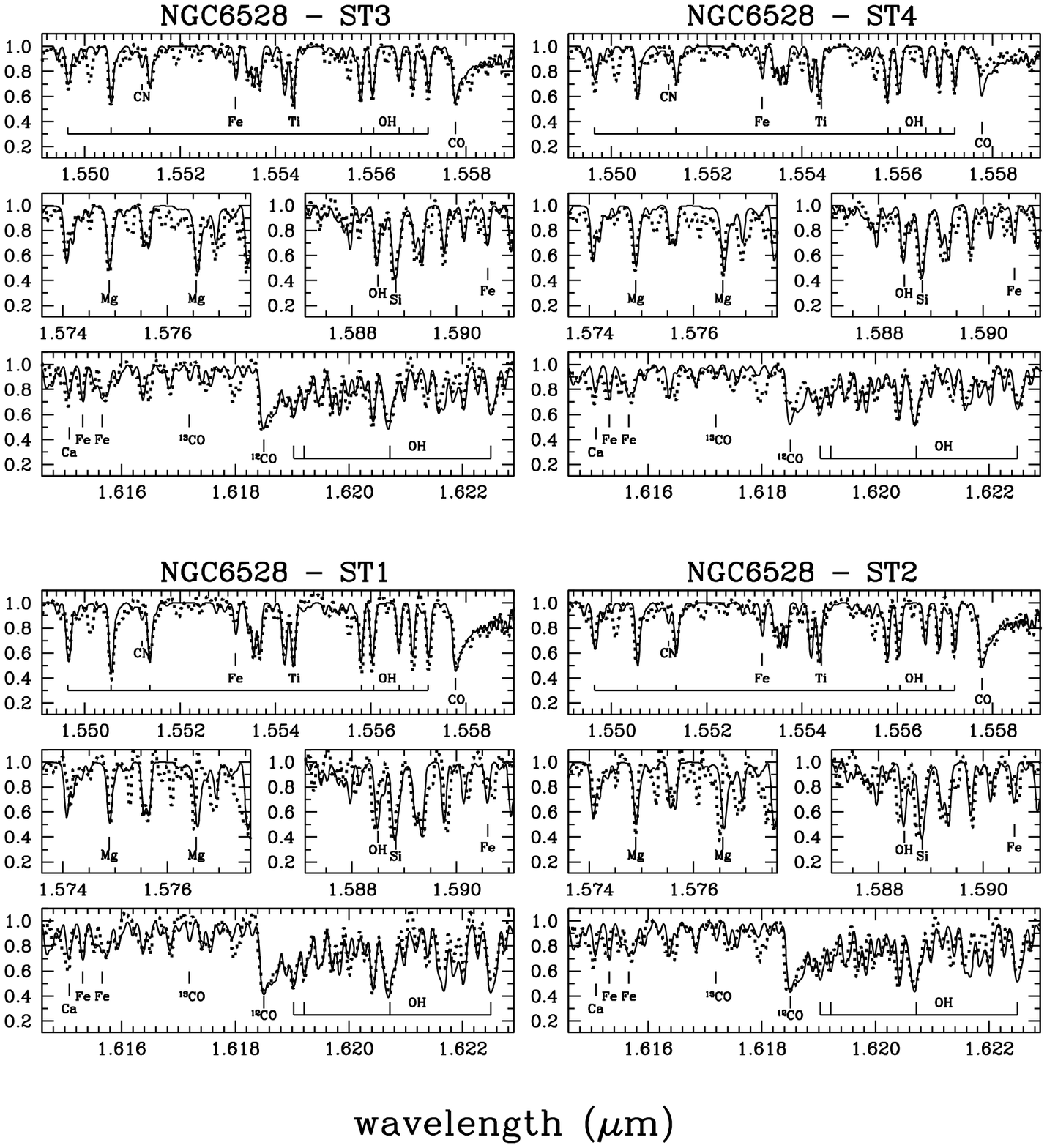}
\caption{
Selected portions of the observed echelle spectra (dotted lines) of the
four giants in NGC~6528 with our best-fitting synthetic spectrum
(solid line) superimposed. A few important molecular and atomic lines
of slightly interest are marked.
}
\label{fig4}
\end{figure*}

Fig.~\ref{fig4} shows
our synthetic best-fitting superimposed on the
observed spectra of the four giants in NGC~6528.
For this cluster our abundance analysis give an average [Fe/H]$=-0.17\pm0.01$,
[O/Fe]$=+0.33\pm0.02$ and an overall average $\rm [\alpha/Fe]=0.33\pm0.01$.
We also measure an average carbon depletion ([C/Fe]=--0.35$\pm0.06$~dex),
a low $^{12}$C/$^{13}$C$\approx8\pm2$ and some [N/Fe] enhancement 
(by a factor of 2-3).

In order to further check the robustness of our best-fitting solutions,  
the same statistical test done for NGC~6342 has been repeated here.
As shown in Fig.~\ref{fig5} our best-fitting solutions  
have an average probability $P>$0.97 to be statistically representative 
of the observed spectra, while the other {\it test models} are only 
significative at $\ge1.5~\sigma$ level. 

By measuring the stellar radial velocities (see Table~\ref{tab1})
we find an average value of $+210\pm $4~km/s with a dispersion $\sigma=7$~km/s, 
in excellent agreement with the one proposed by \citet{car01}, namely 
$\rm v_r=+210\pm2$~km/s and $\sigma$=4~km/s.

\section{Discussion and Conclusions}

Our high resolution spectroscopy in the near IR gives iron abundances for 
NGC~6342 and NGC~6528 which are in excellent agreement 
with the photometric
estimates, as obtained from the RGB morphology and luminosity in the optical 
as well as in the near IR. 

Our iron abundance of NGC~6528 is also in excellent agreement with the  
one estimated by \citet{zoc04} from optical spectroscopy, while is slightly 
lower ($\approx$0.2 dex) than that of  \citet{car01}.
Such a slightly higher value of iron abundance is still marginally consistent 
at a $\ge 2\sigma$ level with our IR spectra. 
A similar iron abundance discrepancy between IR \citep{ori02,mel03} 
and optical measurements by \citet{car01} has been found for NGC~6553.

Our O, Ca, Mg and Ti abundances of NGC~6528 are fully consistent 
with the study of red HB stars by
\citet{car01}, while our Si abundance is lower by $\approx$0.3 dex.
Our O, Si, and Mg abundances are also consistent (only marginally higher)  
with those for red giants (though less luminous than ours)
by  \citet{zoc04}. 
The abundances of Ti and Ca, in particular, by \citet{zoc04} 
\citep[see also][]{mr04} are significantly lower than those from our and 
\citet{car01} studies. 

Our composition is also consistent with that found for Galactic bulge
field stars \citep{mr94}.  
Photometry of NGC 6528 and a comparably metal rich bulge globular cluster,
NGC 6553, indicates that these two clusters are as old as the Galactic
halo with no measurable age difference \citep{ort95}.  This has been reinforced
by recent photometry in which the cluster members have been separated from
the field by their relative proper motions \citep{zoc01,felt02}.
The results of our study are
consistent with NGC 6342 and 6528 being related to the field population of the
Galactic bulge, their composition bearing the imprint of an early era of 
enrichment by massive star supernovae. 
We find it difficult to entertain the notion that
type I SNe contributed significantly to the enrichment of NGC 6528, or that the
composition of this cluster is different from the other bulge clusters we have
studied.  However, the significant dispersion in the composition of NGC6528,
even based on analyses using high S/N spectra from 8-10m class telescopes,
emphasizes that the composition of metal rich stars is far from being a solved
problem.

The low $\rm^{12}C/^{13}C$ abundance ratios measured in NGC~6342 and NGC~6528
are similar to those measured in bright giants of halo 
\citep{ss91,she96,gra00,vws02,smi02,ori03}
as well as other bulge \citep{ori02,she03,ori04}
globular clusters, over the entire range of metallicities between 
one hundredth and solar.
They can be explained by additional mixing mechanisms due to 
{\it cool bottom processing} in the stellar interiors during the evolution along
the RGB \citep[see e.g.][]{cha95,dw96,csb98,bs99}.

\begin{figure}
\centering
\includegraphics[width=8.7cm]{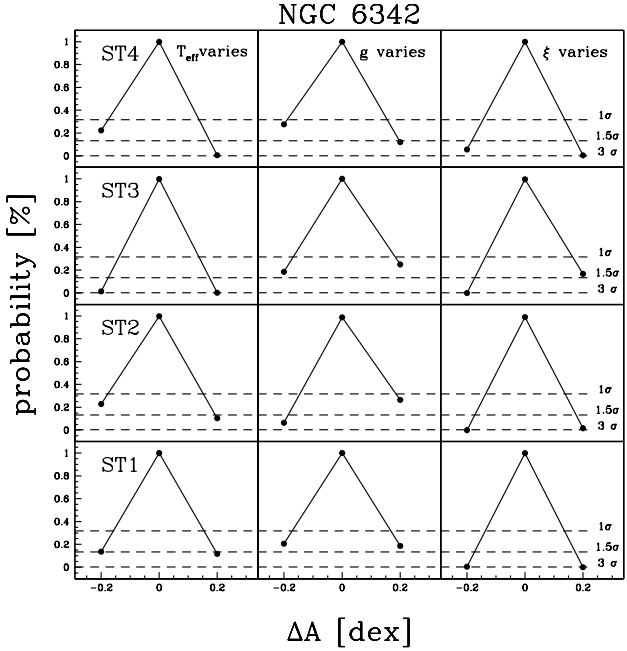}
\caption{
Probability of a random realization
of our best-fitting solution and the test models with varying temperature by 
$\Delta T_{eff}$ of $\pm$200K (left panels), 
gravity by $\Delta log~g$ of $\pm$0.5 dex (middle panels), 
and microturbulence by $\Delta \xi$ of $\mp$0.5 Km s$^{-1}$ (right panels),  
with respect to the best-fitting (see Sect.~\ref{results}) for the four giants 
in NGC~6342.
}
\label{fig3}
\end{figure}

\begin{figure}
\centering
\includegraphics[width=8.7cm]{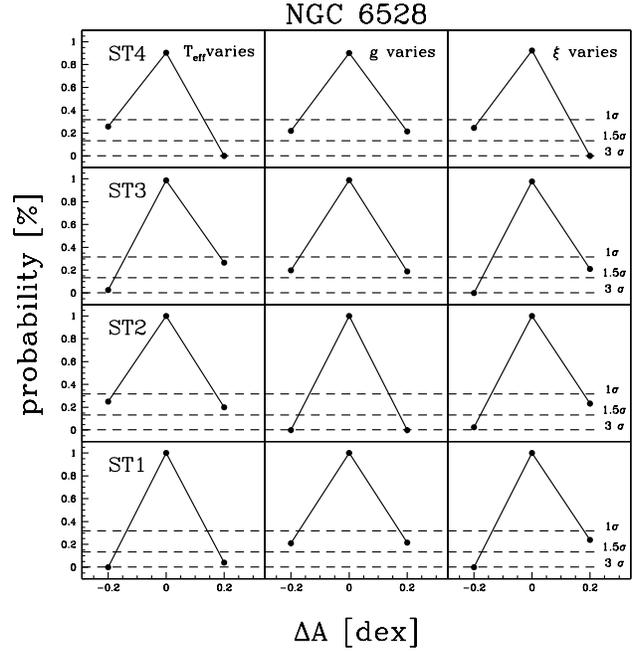}
\caption{
Probability of a random realization
of our best-fitting solution and the test models with varying temperature by
$\Delta T_{eff}$ of $\pm$200K (left panels),
gravity by $\Delta log~g$ of $\pm$0.5 dex (middle panels),
and microturbulence by $\Delta \xi$ of $\mp$0.5 Km s$^{-1}$ (right panels),
with respect to the best-fitting (see Sect.~\ref{results}) for the four giants
in NGC~6528.
}
\label{fig5}
\end{figure}


\section*{Acknowledgments}

LO and EV acknowledge the financial support by the Agenzia 
Spa\-zia\-le Ita\-lia\-na
(ASI) and the Ministero dell'Istru\-zio\-ne, Universit\`a e Ricerca (MIUR).

EV acknowledges the financial support by the {\it Marco Polo} project
and UCLA for the hospitality during her visit.

RMR acknowledges support from grant number AST-0098739,
from the National Science Foundation.
The authors are grateful to the staff
at the Keck Observatory and to Ian McLean
and the NIRSPEC team.
The authors wish to recognize and acknowledge the very significant cultural
role and reverence that the summit of Mauna Kea has always had within
the indigenous Hawaiian community.
We are most fortunate to have the opportunity to conduct observations 
from this mountain.

\label{lastpage}

\end{document}